\documentclass[bookmarks=false]{article}
\usepackage{arxiv}

\usepackage[utf8]{inputenc}
\usepackage[T1]{fontenc}
\usepackage{url}
\usepackage{booktabs}
\usepackage{amsfonts}
\usepackage{nicefrac}
\usepackage{microtype}
\usepackage{graphicx}
\usepackage{multirow}
\usepackage{array}
\usepackage{amsmath}
\usepackage{amssymb}
\usepackage{doi}
\usepackage{authblk}
\usepackage{algorithmic}
\usepackage{algorithm}
\usepackage{cleveref}
\usepackage{float}
\usepackage[numbers,sort&compress]{natbib}
\setcitestyle{square} % or [square]
\usepackage{hyperref}

\title{Enhancing Trust in AI Marketplaces: Evaluating On-Chain Verification of Personalized AI Models using zk-SNARKs}

\author[1]{Nishant Jagannath\thanks{\texttt{Nishant.Jagannath@canberra.edu.au}}}
\author[1]{Christopher Wong}
\author[2]{Braden McGrath}
\author[1]{MD Farhad Hossain}
\author[1]{Asuquo A. Okon}
\author[3]{Abbas Jamalipour}
\author[1]{Kumudu S. Munasinghe}

\affil[1]{School of IT and Systems, University of Canberra, ACT, Australia}
\affil[2]{School of Engineering and Technology, University of New South Wales, ACT, Australia}
\affil[3]{School of Electrical and Information Engineering, University of Sydney, NSW, Australia}

\renewcommand{\shorttitle}{Enhancing Trust in AI Marketplaces on Blockchain with zk-SNARKs}

\hypersetup{
  pdftitle={Enhancing Trust in AI Marketplaces: Evaluating On-Chain Verification of Personalized AI Models using zk-SNARKs},
  pdfsubject={cs.CR, cs.AI, cs.DC},
  pdfauthor={Nishant Jagannath, Christopher Wong, Braden McGrath, MD Farhad Hossain, Asuquo A. Okon, Abbas Jamalipour, Kumudu S. Munasinghe},
  pdfkeywords={AI Verification, zk-SNARKs, Blockchain, Chainlink, Personalized Models, On-Chain AI},
}

\begin{document}

\maketitle

\begin{abstract}
The rapid advancement of artificial intelligence (AI) has brought about sophisticated models capable of various tasks ranging from image recognition to natural language processing. As these models continue to grow in complexity, ensuring their trustworthiness and transparency becomes critical, particularly in decentralized environments where traditional trust mechanisms are absent. This paper addresses the challenge of verifying personalized AI models in such environments, focusing on their integrity and privacy. We propose a novel framework that integrates zero-knowledge succinct non-interactive arguments of knowledge (zk-SNARKs) with Chainlink decentralized oracles to verify AI model performance claims on blockchain platforms. Our key contribution lies in integrating zk-SNARKs with Chainlink oracles to securely fetch and verify external data to enable trustless verification of AI models on a blockchain. Our approach addresses the limitations of using unverified external data for AI verification on the blockchain while preserving sensitive information of AI models and enhancing transparency. We demonstrate our methodology with a linear regression model predicting Bitcoin prices using on-chain data verified on the Sepolia testnet. Our results indicate the framework’s efficacy, with key metrics including proof generation taking an average of 233.63 seconds and verification time of 61.50 seconds. This research paves the way for transparent and trustless verification processes in blockchain-enabled AI ecosystems, addressing key challenges such as model integrity and model privacy protection. The proposed framework, while exemplified with linear regression, is designed for broader applicability across more complex AI models, setting the stage for future advancements in transparent AI verification.
\end{abstract}

\section{Introduction}
\label{sec:introduction}
The proliferation of artificial intelligence (AI) has revolutionized the digital landscape, driving a growing demand for personalized, efficient, and reliable AI models. Developing such models, however, is resource-intensive and requires specialized expertise \cite{hao2019computing}. To bridge this gap, AI marketplaces have emerged as pivotal platforms that facilitate the exchange of personalized AI services. These marketplaces empower developers to monetize their models, providing access to sophisticated AI tools for users who may lack the capacity to develop them independently. A prime example of this trend is the ChatGPT Store \cite{openai2024gptstore}, which offers diverse AI models tailored to various user needs. By enabling the buying, selling, and sharing of pre-trained AI models, AI marketplaces function much like software app stores but with a focus on AI capabilities rather than applications.

\begin{figure}[htpb!]
\centering
\includegraphics [width=7in,height= 7in,keepaspectratio]{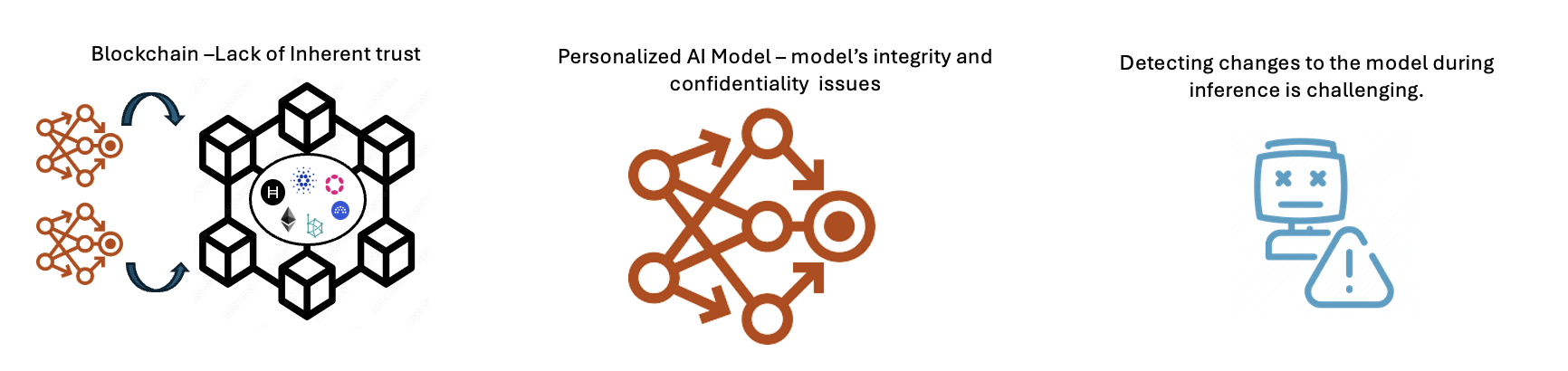}
\caption{The importance of personalized AI model verification on blockchain. 
\label{fig:1}}
\end{figure}

Despite the promise of AI marketplaces, the dominance of a few global tech giants in AI technology has raised significant concerns regarding transparency, fairness, and equitable access \cite{7163223}. Model weights are essential for providing experimental reproducibility and fostering innovation. The push towards commercializing AI models has led to a trend of closed-source models, keeping model weights and other details confidential. This confidentiality is due to the significant investments in data acquisition, computational resources, and algorithmic optimization. Even if developers wish to substantiate the performance claims of their models, publishing these weights could result in the misuse of AI models, leading to advanced cyberattacks or the propagation of disinformation \cite{RAND}. These limitations hinder the examination of model performance and the verification of any claims regarding their effectiveness.

The problem is exacerbated in AI marketplaces operating in decentralized settings, such as blockchain, where there is no inherent trust among users \cite{Sarpatwar2019BlockchainEA}. This lack of transparency makes it difficult to identify performance characteristics, such as performance claims, in production AI models. Ensuring the integrity and reliability of personalized AI models in these marketplaces is crucial, as providers must guarantee model performance, and consumers seek assurance of quality and value. Currently, methods like SingularityNET's decentralized reputation system rely on community participation to rate AI services \cite{goertzel2017singularitynet}. However, this method lacks the rigour necessary for comprehensive validation. These issues as seen in  Fig.~\ref{fig:1}, highlight the need for a decentralized and transparent verification mechanism that fosters trust.

\begin{figure}[h!]
\centering
\includegraphics[width=7in,height= 7in,keepaspectratio]{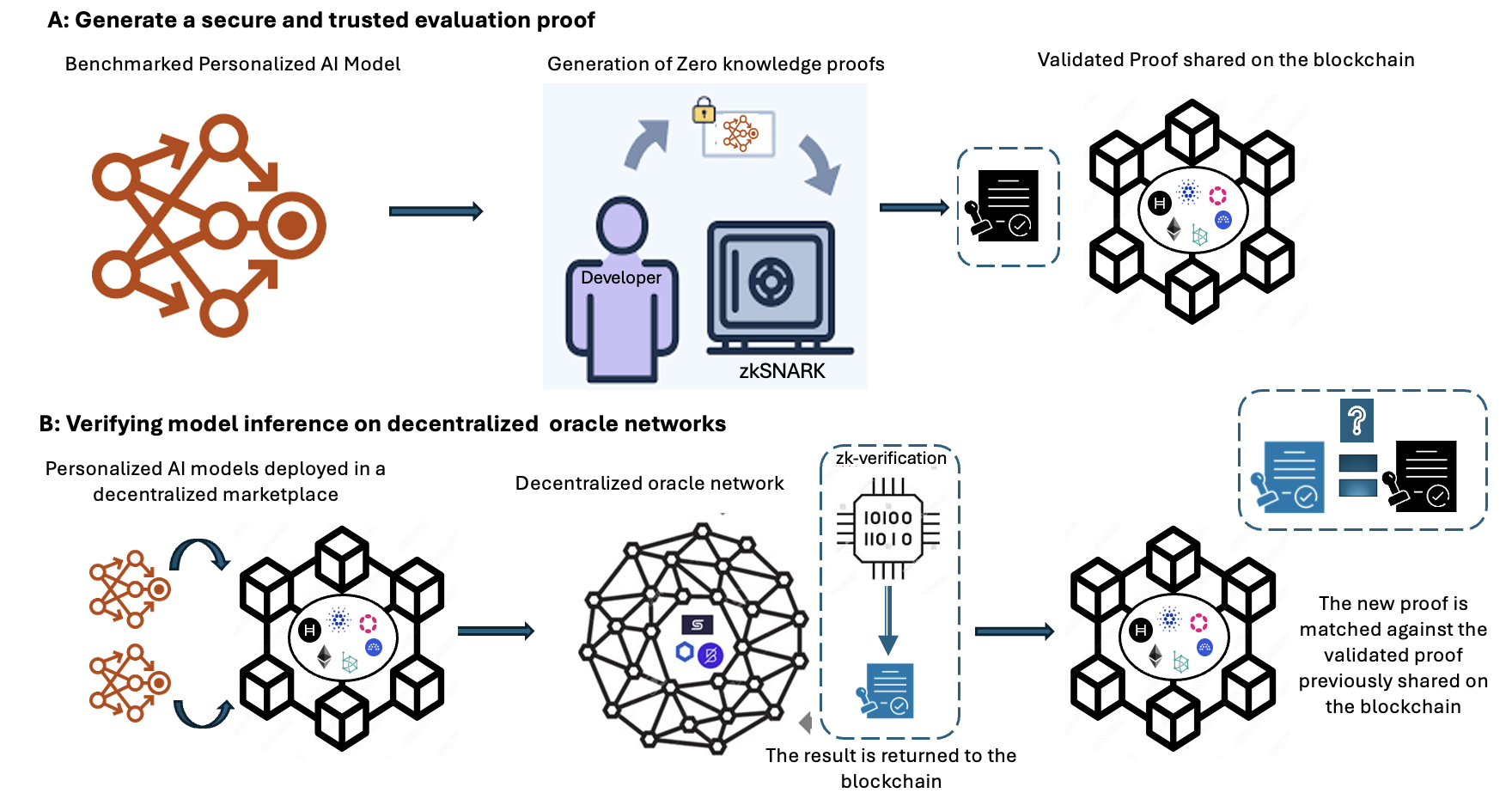}
\caption{A high-level overview of the system design. 
\label{fig:2}}
\end{figure}

Technologies like Zero-Knowledge Succinct Non-Interactive Arguments of Knowledge (zk-SNARKs) can help address trust and model privacy issues in this context. zk-SNARKs provide powerful cryptographic proofs that verify the correctness of computations without revealing the underlying data \cite{chen2023reviewzksnarks}. However, using zk-SNARKs to verify AI models' integrity and performance claims on blockchain-based marketplaces presents several challenges. Firstly, the compactness of zk-SNARK verification proofs is offset by the substantial resources needed for proof verification, potentially causing bottlenecks \cite{Garoffolo2024SnarktorAD}, especially when the blockchain handles multiple transactions and interactions simultaneously. Secondly, the computational intensity of zk-SNARK proofs involves complex mathematical computations that are both time-consuming and costly in terms of blockchain gas fees on platforms like Ethereum \cite{Garoffolo2024SnarktorAD}. Furthermore, verifying claims of AI models using zk-SNARKs often requires external data inaccessible within the blockchain \cite{Sarpatwar2018TowardsET}. These considerations highlight the need for a decentralized approach that leverages off-chain computation for data collection and verification, and on-chain verification to optimize the performance and scalability of blockchain-based AI marketplaces.

Decentralized oracles are critical in bridging blockchain technology with the external world to validate transactions. They present untapped potential for verifying AI models in marketplaces. By bridging the digital and physical realms, oracles can conduct rigorous assessments of AI models' claims, ensuring they meet high standards before being made available. This paper explores a novel approach as shown in Fig.~\ref{fig:2} that integrates zk-SNARKS on Chainlink's \cite{ChainlinkWhitepaperV2} decentralized oracle network with blockchain to verify AI models. This approach could revolutionize the development and distribution of personalized AI services by enhancing trust in blockchain-enabled AI marketplaces. This approach has practical applications across various sectors that require verifiable computation, such as finance, healthcare, education and supply chain management, where accurate AI model predictions are critical and transparency is paramount. 
By implementing such a solution, we can create a more open, equitable, and reliable AI marketplace, driving the next wave of advancements in AI technology.

\subsection{Contributions}
This paper addresses the challenges of secure and efficient verification of personalized AI models in a blockchain-enabled AI marketplace. We present a comprehensive study using zk-SNARKs and Chainlink oracles. The key contributions of this paper are as follows:
\begin {itemize}
\item A novel comprehensive framework that leverages decentralized oracles (Chainlink) to validate unverified data from off-chain data sources for zk-SNARK proof verification, ensuring transparent and trustless verification of AI models on blockchain while preserving model privacy.

\item A working implementation that integrates zk-SNARKs with Chainlink oracles, demonstrating their practical use in AI model verification scenarios.

\item Analysis of the efficiency and resource consumption of zk-SNARK proof generation and verification to identify key areas for optimization.

\item Analysis of the computational costs such as transaction fees and LINK token costs associated with zk-SNARK verification's, providing insights into the costs involved.
\end{itemize}

This article is organised as follows. Section II provides an overview of relevant work, emphasising current research on verification of AI models in decentralized settings. Section III covers the system architecture and is divided into four subsections: A, B and C. Subsection A describes the method used to generate a secure and evaluation proof. Subsection B describes the method used to verify model inference  while subsection C provides an overview of the proposed framework, D describes the proposed system model. Section IV describes the experimental setup, whereas Section V presents the results and their interpretation. Finally, in Section VI, we summarise our findings and conclusions and outline areas for further research.

\section{Literature Review}
Recent advances in AI models have led to significant progress in various decentralized systems, particularly in the integration of AI with blockchain technology. This development has huge potential for revolutionizing various industries and domains \cite{salah}. The benefits of this integration as highlighted by \cite{chavali} and \cite{montes} include improved system performance and a more equitable development of AI. Furthermore, various techniques and applications of decentralized AI, such as decentralized machine learning (ML) frameworks and distributed AI marketplaces are explored in \cite{vincent}.

Traditional trust mechanisms for ensuring the trustworthiness of AI models have been extensively researched, with various approaches proposed. Key issues include transparency and interpretability \cite{upreti}, robustness and fairness \cite{mylrea}, uncertainty quantification \cite{abdar}, and causal reasoning \cite{pearl}. Transparency and interpretability are crucial for building trust in AI models and making the decision-making process understandable to humans. Techniques such as model visualization, saliency maps, Local Interpretable Model-agnostic Explanations (LIME), and SHapley Additive exPlanations (SHAP) support these goals \cite{upreti}. Robustness and fairness are also vital components of trustworthy AI systems, with techniques like adversarial training and data augmentation enhancing robustness against attacks, while debiasing algorithms and fairness constraints mitigate discriminatory biases \cite{mylrea}. Uncertainty quantification, using methods such as Bayesian neural networks, ensemble methods, and conformal prediction, provides a measure of confidence in AI model predictions, particularly important in critical domains such as healthcare and autonomous systems \cite{abdar}. Causal reasoning, facilitated by tools such as causal inference, structural causal models, and counterfactual reasoning, is essential for achieving a more interpretable and robust decision-making framework in AI models \cite{pearl}.  

Despite these multi-faceted strategies for developing trustworthy centralized AI systems, traditional trust mechanisms often fail to preserve data privacy and confidentiality in decentralized systems, where data is replicated across multiple nodes. In addition, decentralized systems face scalability and performance limitations, making it challenging to handle large-scale applications and high transaction volumes using traditional centralised approaches. Major challenges of traditional trust mechanisms in decentralized environments include the lack of a central authority, identity verification issues, Sybil attacks, scalability and consistency issues, and legal and regulatory uncertainty \cite {arshad}, \cite{filippi}, \cite{majdoubi}.

The landscape of AI has entered a new era with the advent of blockchain-enabled AI marketplaces. These marketplaces enable individuals and organisations to decentralise AI models' sharing, trading, and utilisation, in a manner that democratises access to advanced AI technologies \cite{goertzel2017singularitynet}. Despite their numerous benefits, decentralized marketplaces present unique challenges for authenticating and verifying AI models. The diversity and volume of AI models exchanged on these platforms render traditional centralised verification and validation processes impractical. Consequently, there is an urgent need for novel approaches to perform these crucial functions efficiently and dependably. The Neuromation platform is an AI marketplace that leverages synthetic data for training models, substantially reducing the time and cost associated with developing AI models. Additionally, they possess a distributed computing platform designed for model training

Chainlink is a pioneering decentralized oracle network that seamlessly connects smart contracts on blockchains with off-chain data and systems \cite{ChainlinkWhitepaperV2}. As a secure middleware, it enables blockchain applications to reliably access and leverage real-world information, unlocking a vast array of innovative use cases. At its core, Chainlink employs a decentralized network of independent oracle nodes that retrieve and deliver data to smart contracts, mitigating single points of failure \cite{ezzat}. Through crypto-economic incentives and penalties, it ensures the reliability and correctness of oracles, even against well-resourced adversaries.

Chainlink enhances blockchain scalability and efficiency by enabling secure off-chain computations and data processing, which are then integrated on-chain, facilitating the development of advanced hybrid smart contracts \cite{ezzat}. Through its confidentiality measures and trust minimization achieved via decentralization and cryptographic assurances, Chainlink acts as a secure conduit between blockchains and real-world data, driving the evolution and broader adoption of sophisticated decentralized applications across various sectors \cite{zhao}. Recent research suggests that the integration of AI and blockchain could be further enhanced with Chainlink \cite{bhumichai}, which ensures the integrity and transparency of data inputs used in AI models, thereby providing a robust foundation for the ethical and verifiable deployment of AI technologies.

Verification and validation (V\&V) are necessary quality assurance procedures for preserving the trust and dependability of AI systems. Verification ensures that the AI model was implemented accurately and behaved as intended per its mathematical description \cite{oberkampf_roy_2010}. It is comparable to "building the model properly." Validation, conversely, guarantees that the AI model satisfies the requirements of the context or problem it was designed to solve – it involves "building the right model". Despite their robust capabilities, AI models occasionally generate inaccurate predictions and manifest unintended behaviour. 

These risks may be exacerbated in high-stakes domains such as healthcare or finance, where errors may result in severe adverse outcomes, from incorrect medical diagnoses to substantial financial losses. This makes V\&V processes essential for the safety and dependability of AI systems, assuring that their decisions are accurate, trustworthy, and dependable \cite{rudin2019stop}. As these models take on increasingly complex duties, their verification and validation become paramount \cite{amodei2016concrete}. These procedures are essential for maintaining confidence in AI systems because they help identify and mitigate risks associated with inaccurate predictions or biased outcomes \cite{rudin2019stop}.

A key challenge in AI is verifying personalized, closed-source models in a way that safeguards sensitive information, preserves intellectual property, and enhances transparency, as traditional methods often rely on trust or costly re-evaluation. To this end, zero-knowledge proofs have emerged as a powerful tool for privacy-preserving authentication \cite{Goldwasser1985TheKC}. This cryptographic technique allows one party to prove to another that a given statement is true, without revealing any additional information about the statement.  Initially, the zk proofs were designed to be interactive and could not be re-verified multiple times by other validators without creating new interactions. This led to the development of Non-Interactive Zero-knowledge Proofs (NIZKPs) \cite{Blum}, allowing the zero-knowledge proofs to be re-verified by multiple parties.  

There are several popular implementations of zero-knowledge proofs, including zk-SNARKs\cite{inventzksnarks}, Zero-Knowledge Scalable Transparent Argument of Knowledge (zk-STARKs) \cite{zkstarks} and bulletproofs \cite{bulletproofs}. 
One of the primary differences between zk-SNARKs, zk-STARKs and bulletproofs is the trusted setup process. An initial trusted setup process is required for zk-SNARKs and it's not required by zk-STARKs and bulletproofs. zk-STARKs have larger proof sizes, resulting in higher verification costs and storage requirements on the blockchain. Bulletproofs have smaller proof sizes but require interactive verification, which is less practical for decentralized systems. Beyond zero-knowledge proof systems, there exist other cryptographic techniques for verifying computations with privacy guarantees, such as Homomorphic Encryption (HE), Verifiable Computing (VC) \cite{HEVC} and Secure Multiparty Computation (MPC) \cite{Lindell2020SecureMC}. While these methods are widely used for general secure computation and data confidentiality, they are not specifically tailored for AI-based tasks.

For this research, we consider zk-SNARKS, despite their reliance on a trusted setup. zk-SNARKs achieves significantly smaller proof sizes compared to zk-STARKs and bulletproofs, resulting in smaller shorter verification times and less gas cost \cite{Xie2019LibraSZ}. In the context of personalized AI models, zk-SNARKs can be leveraged to verify the correctness of a model's predictions without disclosing the underlying model parameters or training data \cite{chen}. This is particularly relevant when AI models are deployed in environments handling sensitive user data.

The US Department of Energy implemented a secure neural network verification system using zk-SNARKs for Nuclear Treaty Verification \cite{SNNzksNARK}. This proposed system allows to verify the neural network output, input hash and Rivest–Shamir–Adleman (RSA) signature with zk proof, enabling a secure, adaptable way to disclose sensitive data on nuclear materials and facilities. The work by \cite{south} investigates verifiable evaluation attestations using zk-SNARKs, enabling independent validation of model performance claims without exposing the models' internal weights or outputs. Here the authors employ a "predict, then prove" strategy, where models are converted to a standard format, evaluated on benchmark datasets, and proofs of correct inference are generated. These proofs are aggregated into attestations that can be independently verified.  

The authors in \cite{Kang2022ScalingUT} presented a practical approach to verify ML model inference for a full-resolution ImageNet model using zk-SNARKs and explore other scenarios such as verifying MLaaS predictions and accuracy. The zk-SNARKs enabled a non-interactive way to verify ML model execution and achieved 79\% accuracy. A scheme called zkCNN was proposed to prove the accuracy of a convolution neural network (CNN)  model's predictions using public dataset to others without revealing sensitive information about the model \cite{Liu2021zkCNNZK}.

Based on our literature review, it is evident that technologies like zk-SNARKs can help address trust and AI model privacy issues in this context \cite{SNNzksNARK}, \cite{south}, \cite{Kang2022ScalingUT}, \cite{Liu2021zkCNNZK}. However, using zk-SNARKs to verify AI models' integrity and performance claims on blockchain-based marketplaces presents several challenges. Verifying claims of AI models using zk-SNARKs often requires external data inaccessible within the blockchain \cite{Sarpatwar2018TowardsET}. Similar to the work in \cite{Liu2021zkCNNZK}, models can be trained on public datasets and to prove the model accuracy claims, access to high quality public datasets are required.  The compactness of zk-SNARK verification proofs is offset by the substantial resources needed for proof verification, potentially causing bottlenecks \cite{Garoffolo2024SnarktorAD}, \cite{excesszksnarks} especially when the blockchain handles multiple transactions and interactions simultaneously. Secondly, the computational intensity of zk-SNARK proofs involves complex mathematical computations that are both time-consuming and costly \cite{Wahby2018DoublyEfficientZW} especially in terms of blockchain gas fees on platforms like Ethereum \cite{Garoffolo2024SnarktorAD}. These considerations highlight the need for a decentralized approach that leverages off-chain computation for data collection and verification and on-chain zk verification to optimize the performance, scalability and enhancing trust within blockchain-based AI marketplaces.

This paper addresses existing gaps by proposing a novel framework that leverages zk-SNARKs integrated with Chainlink oracles to verify AI model performance claims on blockchain platforms. Our approach allows for the verification of personalized AI models without disclosing sensitive information, preserving intellectual property and enhancing transparency. We demonstrate our approach with a linear regression model predicting Bitcoin prices using on-chain data, verified on the Sepolia testnet.

\begin{figure}
    \centering
    \includegraphics[width=1\linewidth]{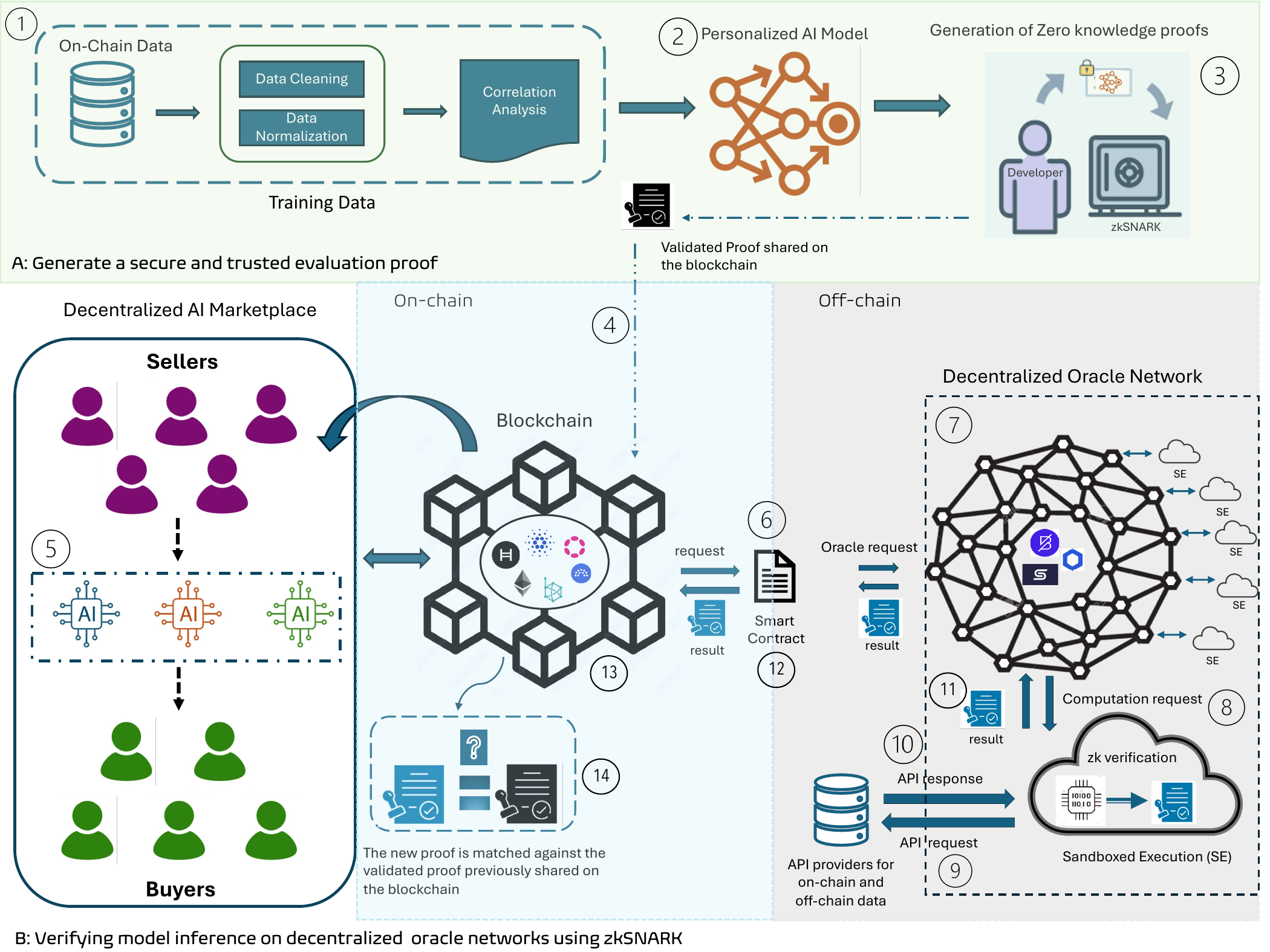}
    \caption{Proposed verification framework.}
    \label{fig:3}
\end{figure}

\section{Methodology and System Design}
This section describes the methodology as shown in Fig.~\ref{fig:3} for verifying the performance claims of a personalized AI model without revealing weights and are trained on on-chain and user-specific data to predict Bitcoin prices. The verification process is computed on Chainlink's decentralised oracle network using zk-SNARKs. We divide the section into two parts and explain these parts with respect to Fig.~\ref{fig:3}. In Part A, we provide the system overview of our proposed framework. Part B outlines the steps to generate a secure and trusted evaluation proof and Part C describes the verification process for the model inference on a decentralized oracle network using zk-SNARKs.

\subsection{System Overview}
Trust is a major concern for users in the Web3 domain, particularly on the blockchain. Trust issues also extend to the blockchain-enabled AI marketplace, where the credibility of developers' performance claims for personalized AI models is questioned. The blockchain-enabled AI marketplace combines on-chain and off-chain elements to enhance the verifiability of verifications. The framework represented in Fig.~\ref{fig:3} is specifically designed to enable personalized AI model performance verification using zk-SNARKs. The interaction between on-chain smart contracts and off-chain Chainlink oracles is crucial for the functioning of the blockchain-enabled AI marketplace. The interaction guarantees that the data, computation, and proof validation are carried out securely and efficiently. We analyze the on-chain data from external API providers and eliminate inaccurate data points. The data is carefully scaled to uncover and examine the connections between important data points in the model's output. After the training and testing of the personalized AI model, developers generate zk-SNARK proofs to verify the AI model's claim without exposing sensitive data such as model weights. These verifiable proofs are shared on the blockchain. 

Prior to purchasing the personalized AI model, the buyer demands proof to verify the performance claim of the personalized AI model. The decentralized oracle network is used for verification using the Chainlink Functions, as requested by the blockchain. The Chainlink nodes facilitates the coordination of data acquisition from external API providers for on-chain data and the execution of computations. Each node in the Chainlink carries out sandboxed execution of the provided source code to ensure transparency. The aggregated results are sent to the smart contract using Chainlink's Off-Chain Reporting (OCR) protocol \cite{chainlink2024ocr}. The smart contract on Sepolia receives the aggregated result and zk-SNARK proof. 

The blockchain verifies the proof using stored verification keys and updates the state of the blockchain based on the verification outcome. Verified proofs are stored on-chain for future reference, ensuring a transparent and tamper-proof record of all computations. This framework provides a practical approach to verifying personalized AI models. Incorporating zk-SNARK ensures the privacy of model weights during verification, enhancing trust and transparency in AI model marketplaces. The integration of zk-SNARKs into Chainlink functions facilitates secure and reliable data fetching and computation, offering a robust AI model verification framework that can be implemented in real-world scenarios.

To summarize the interactions in the proposed verification framework, Fig.~\ref{fig:3} represents the framework for verifying personalized AI model performance using zk-SNARKs. In Step 1, The process begins with developers training personalized AI models, followed by data cleaning, normalization, and correlation analysis. In Step 2, developers generate zk-SNARK proofs to verify model performance claims without revealing sensitive data and upload these proofs to the blockchain. In Step 3, buyers initiate verification requests, which the decentralized oracle network processes by fetching data from external APIs and performing zk-SNARK verification in a sandboxed environment. In Step 4, the Chainlink oracles communicates results back to the blockchain smart contract via Chainlink’s Off-Chain Reporting protocol. In  Step 5, the blockchain then validates the proofs using stored verification keys and updates the state of the decentralized marketplace, ensuring a transparent and tamper-proof record of all zk-SNARK verifications. 

\subsection{Generate a Secure and Evaluation Proof}

\subsubsection{Personalized AI model - Introduction to Personalization}

Personalized AI models provide customized predictions by utilizing on-chain data and user data. For example, the model can be personalized when predicting Bitcoin prices to consider the user's unique trading patterns, preferences, and other data points affecting their investment choices.

\paragraph{Data Collection:}
Developers acquire on-chain data in two ways. The first method involves collecting and processing raw on-chain data from the public Bitcoin blockchain. The second method uses external application programming interface (API) providers where the on-chain data is already preprocessed and ready to use. We obtained on-chain data from 2016 to 2023 from API providers such as \cite{block_2021}, \cite{glassnode_studio1}. With the on-chain data collected from these sources, we categorized and analyzed metrics from each category against Bitcoin's price. We also use user-specific data such as transaction history and wallet activity. The following metrics are obtained from the on-chain data; block size, block height, transaction count, daily active addresses, miners revenue, miner fees, miner to exchanges,  total new addresses, transactions rate, transfers count, hash rate, transactions difficulty, transfer rate, wallets address with greater than 1, 10 and 100 coins, exchange deposits, exchange withdrawals and total addresses.

\paragraph{Data Analysis:} The on-chain data closely correlating to the bitcoin price are identified. This step involves using a Pearson and Spearman correlation analysis to understand the linear and non-linear relationship between the on-chain datasets and the bitcoin price. The Pearson correlation can be represented by equation (\ref{eq:1})

\begin{equation} \label{eq:1}
r = \frac{\sum_{i=1}^n (x_i-\bar{x})(y_i-\bar{y})}{\sqrt{\sum_{i=1}^n(x_i-\bar{x})^2\sum_{i=1}^n(y_i-\bar{y})^2}}
\end{equation}

where:
\begin{itemize}
    \item $x_i$ is the $i$-th data point of features of on-chain data    
    \item $y_i$ is the $i$-th data point of bitcoin price
    \item $\bar{x}$ is the mean of the $x$ values
    \item $\bar{y}$ is the mean of $y$ values
    \item $n$ is the total number of data points
\end{itemize}

The Spearman rank correlation coefficient \cite{Spearman} is a nonparametric measurement correlation used to evaluate the monotonic relationship between two variables.

\begin{equation}
    \rho = 1 - \frac{6\sum {d_i}^2}{n({n}^2 -1)}
\end{equation}

In (2), the difference between the ranks of the $i$-th pair of values is represented by $d_{i}$ and $n$ represents the total number of data points.

Pearson correlation coefficients are used to quantify the linear connection between variables. In contrast, Spearman correlation coefficients are only applicable to monotonic connections, in which variables tend to move in the same or opposite direction but not necessarily at the same rate. In a linear relationship, the rate is constant.

We re-scale the data between the range [0,1]. The normalization value is calculated using equation (\ref{eq:2}).

\begin{equation} \label{eq:2}
z = \frac {x_i - \min(x)} {\max(x) - \min(x)}
\end{equation}

By conducting correlation analysis, we can pinpoint important on-chain metrics that can be incorporated into advanced predictive algorithms. Conversely, we can also identify metrics that could be more relevant and should be considered.

\paragraph{Introduction to zk-SNARKs:}
A zk-SNARK allows a prover to convince a verifier that they know a solution to a computational problem without disclosing the solution itself. These proofs are short and fast to verify, and they do not require ongoing interaction between the prover and the verifier after the initial setup. A zk-SNARK system comprises three core algorithms: Generation (Gen), Prover (P) and Verification (V).

\paragraph{Non-Interactive Zero-Knowledge Argument}
The arithmetic circuits in zk-SNARKs play a critical role in representing the computational problem that the prover aims to demonstrate to the verifier that it has been solved correctly. In the context of the non-interactive zero-knowledge argument, let $C$ be an arithmetic circuit such that $C: F^n \times F^{n'} \rightarrow F^l$. Here, $F$ denotes a finite field and $F^{n}$ represents a vector space of dimension $n$ over the finite field $F$. Similarly, \(F^{n'}\) and \(F^l\) indicate vector spaces of dimensions \(n'\) and \(l\) over \(F\), respectively. The NP language $L$ is defined as the set of statements $x$ in $F^n$ for which there exists a valid witness $w$ in $F^{n'}$. This is represented by the relation $R$ defined as $R := \{(x, w) \in F^n \times F^{n'}\}$, where $w$ is the witness and $x$ is the statement.

A non-interactive zero-knowledge argument for the relation $R$ consists of the triple of polynomial-time algorithms: Generation (Gen), Prover (P), and Verification (V).
\begin{itemize}
    \item Generation (Gen): Produces a common reference string (crs) and a private verification state.
    \[
    (\text{crs}) \leftarrow \text{Gen}(1^n, R)
    \]
    \item Prover (P): Produces a proof $\pi$ for a statement $x$ using a witness $w$.
    \[
    \pi \leftarrow \text{P}(\text{crs}, x, w)
    \]
    \item Verification (V): Verifies the proof $\pi$ for the statement $x$.
    \[
    \text{V}(\text{crs}, x, \pi) \rightarrow \{0, 1\}
    \]
\end{itemize}

\paragraph{Properties of zk-SNARKs}
The following properties \cite{Gennarozk} must be met by a non-interactive zero-knowledge proof $\pi$ for the relation $R$:

\begin{itemize}
    \item \textbf{Completeness}: For a statement $x \in F^n$ with a witness $w \in F^{n'}$ such that $(x, w) \in R$, the prover acting honestly always produces a valid proof $\pi$. This proof should be sufficient to convince an honest verifier. The completeness of the non-interactive zero-knowledge proof can be expressed as follows \cite{rodriguez2020}:
    
    \begin{equation}
    \Pr \left[ \begin{array}{c}
    (\text{crs}) \leftarrow \text{Gen}(1^n, R) \\
    \pi \leftarrow \text{P}(\text{crs}, x, w) \\
    \text{V}(\text{crs}, x, \pi) = 1 \text{ if } (x, w) \in R
    \end{array} \right] = 1
    \end{equation}
    
\item \textbf{Soundness}: When an adversary attempts to deceive by providing a proof $\pi$ for a false statement $x \notin R$, the verification algorithm $\text{V}$ is designed to have a high probability of rejecting the proof.  Any evidence \( \pi \) offered by an adversary will be rejected with a high probability due to the soundness requirement, which ensures that \( x \) must be in the relation \( R \) \cite{Goldwasser1985TheKC}:
    
    \begin{equation}
    \Pr \left[ \begin{array}{c}
    (\text{crs}) \leftarrow \text{Gen}(1^n, R) \\
    (x, \pi) \leftarrow \mathcal{A}(\text{crs}) \\
    \text{V}(\text{crs}, x, \pi) = 1 \text{ and } (x, w) \notin R
    \end{array} \right] \leq \text{negl}(n)
    \end{equation}
    
Furthermore, suppose there is an extractor $\mathcal{E}$ that can generate the witness $w \leftarrow \mathcal{E}_{\mathcal{A}}(\text{crs})$ based on the output of an adversary $\mathcal{A}$, which produces a valid argument $(x, \pi) \leftarrow \mathcal{A}(\text{crs})$:

\begin{equation}
    \Pr \left[ \begin{array}{c}
    (\text{crs}) \leftarrow \text{Gen}(1^n, R) \\
    (x, \pi) \leftarrow \mathcal{A}(\text{crs}) \\
    w \leftarrow \mathcal{E}_{\mathcal{A}}(\text{crs}) \\
    \text{V}(\text{crs}, x, \pi) = 1 \text{ and } (x, w) \notin R
    \end{array} \right] \leq \text{negl}(n)
    \end{equation}
    
\item \textbf{Zero-Knowledge}: This characteristic ensures that the verifier only gains knowledge of the statement's truth. In zk-SNARKs, Tau refers to the trusted setup parameter generated during the initial phase, creating a secure cryptographic environment. The Powers of Tau (PoT) ceremony generates these parameters, which are necessary for generating and verifying zk-SNARK proofs, ensuring privacy. In the Phase 2, the crs is further refined to support the specific zk-SNARK application, introducing additional complexity as it tailors the parameters to the operations of the AI model being verified. Together, PoT and Phase 2 form the backbone of the trusted setup, ensuring a robust and reliable foundation for zk-SNARK operations. Without knowing the witness $w$, the proof or argument $\pi$ for a valid assertion $x$ can be simulated using a polynomial-time procedure known as a simulator. Simulator 1 $(S_1)$ generates a simulated proof based on the crs and the random Tau parameter. This demonstrates that the proof system can function without accessing private data thus maintaining the zero-knowledge property. Simulator 2 $(S_2)$ simulates the zk-SNARK proof using the input, output pair and a random Tau. This confirms that the system can generate valid proofs without revealing sensitive information, completing the zero-knowledge simulation. The zero-knowledgeness can be expressed as follows \cite{rodriguez2020}:

\begin{equation}
\footnotesize
\Pr \left[
  \begin{array}{c}
    (\text{crs}) \leftarrow \text{Gen}(1^n, R) \\
    (x, w) \leftarrow \mathcal{A}(\text{crs}) \\
    \pi \leftarrow \text{P}(\text{crs}, x, w) \\
    \mathcal{A}(\pi) = 1
  \end{array}
\right] = \Pr \left[
  \begin{array}{c}
    (\text{crs}, \tau) \leftarrow S_1(1^n, R) \\
    (x, w) \leftarrow \mathcal{A}(\text{crs}) \\
    \pi \leftarrow S_2(\text{crs}, x, \tau) \\
    \mathcal{A}(\pi) = 1
  \end{array}
\right]
\normalsize
\end{equation}
    
\end{itemize}

\subsubsection{Conversion of Linear Regression Model to zk-SNARK Circuit for Validation}
We use zk-SNARKs to generate verifiable computations on-chain of the model without revealing its weights. The linear regression model is converted into a zk-SNARK circuit to represent the model's internal operations. The following steps are used in converting the linear regression model into a zk-SNARK circuit:

\paragraph{Step 1: Model Representation}
The developer trains the personalized AI model, specifically a linear regression model that predicts Bitcoin prices based on historical on-chain data. The model takes various features (independent variables) from the on-chain data and user-specific data, such as transaction history and wallet activity of the user, and predicts the price (dependent variable) of Bitcoin. The linear regression model is represented using (\ref{linear}):
\begin{equation}\label{linear}
y = a_0 + a_1 x_1 + a_2 x_2 + \ldots + a_n x_n + C
\end{equation}

where:
\begin{itemize}
    \item $y$ is the predicted bitcoin price.
    \item $x_i$ are the features of on-chain and user-specific data.
    \item $a_i$ are the coefficients (weights) learned during training.
    \item $C$ is the intercept.
\end{itemize}

\paragraph{Step 2: Arithmetic Circuit Construction}
The linear regression model equation is converted into an arithmetic circuit to permit proving zk-SNARK based computational statements. Each mathematical operation in the linear regression model is mapped to a multiplication and addition gate in zk-SNARKs. For example, the operation $a_1 x_1$ is handled by multiplication gates and sum $a_0$ + $a_1 x_1$ is handled by addition gates. The final output $y$ is computed by using addition gates adding all terms together. This process transforms the linear regression equation into an arithmetic circuit that is compatible with zk-SNARKs. 

\paragraph{Step 3: QAP Conversion}
The models arithmetic circuit are converted into a QAP, providing a framework for zk-SNARKs to check the correctness of the operations in the arithmetic circuit. A QAP for a function $f$ is defined by three sets of polynomials $\{v_i(x)\}, \{w_i(x)\}, \{y_i(x)\}$ and a target polynomial $t(x)$.

For an arithmetic circuit $C$ with $m$ gates:
\begin{equation}
p(x) = \left( \sum_{i=0}^{m} a_i \cdot v_i(x) \right) \cdot \left( \sum_{i=0}^{m} a_i \cdot w_i(x) \right) - \left( \sum_{i=0}^{m} a_i \cdot y_i(x) \right)
\end{equation}
where $t(x)$ divides $p(x)$ and $a_{i}$ represents the coefficients of the polynomials.

The QAP introduces constraints that must be satisfied to ensure all operations in the arithmetic circuit are represented correctly in zk-SNARK form. The complexity of these QAP constraints increases with larger number of features in the linear regression model. As the complexity of the AI models increases, it will require more number of gates to represent model internal operations, leading to higher computational resources and longer proof generation times.

\paragraph{Step 4: zk-SNARK Proof Generation and Verification}
The prover generates a proof $\pi$ demonstrating they know $\{a_i\}$ satisfying the Quadratic Arithmetic Program (QAP) equations:
\begin{equation}   
\pi = (A, B, C)
\end{equation}
where:
\[
A = \sum_{i=0}^{m} a_i \cdot g^{v_i(s)}, \quad B = \sum_{i=0}^{m} a_i \cdot g^{w_i(s)}, \quad C = \sum_{i=0}^{m} a_i \cdot g^{y_i(s)}
\]

The polynomials ${v_i(s)},{w_i(s)},{y_i(s)}$ represent the QAP for the arithmetic circuit. These polynomials are evaluated at a secret value 
$s$. The components of the zk-SNARK proof are represented by $A,B,C$ and the generator of a cryptographic group by $g$, which is used to generate all the elements of the group through its powers. The verifier checks the proof by ensuring:
\begin{equation}
e(A, B) = e(g, C) \cdot e(g^{t(s)}, g)
\end{equation}

where $e(A, B)$ represents the bilinear pairing function used for verification and $t(s)$ is the target polynomial evaluated at the secret value $s$.

\subsection {Verifying Model Inference on Decentralized Oracle Network Using zk-SNARKs}

In this paper, we use the Chainlink Decentralized Oracle Network (DON), hereafter referred to as Chainlink oracles, to perform off-chain computations and relay data to the blockchain. The blockchain component in our framework is represented by the Sepolia testnet, which serves as a proxy for a production blockchain environment. Chainlink Functions enable smart contracts to access a computing infrastructure that is trust-minimized. Smart contracts can access on-chain and off-chain data from APIs and perform personalized computations. By seamlessly integrating these functions with the Sepolia testnet, we can efficiently execute zero-knowledge (zk) verification computations on chainlink's decentralized  oracle network, ensuring that verified results are returned to the blockchain.

Smart contracts utilize the Chainlink nodes to retrieve data from external APIs by sending requests for source code. Every node in the Chainlink carries out the code within a secure and sandboxed execution, efficiently handling the required computations. The zk-SNARK circuits use the obtained data to perform computations without disclosing confidential details. The process yields zk-SNARK proofs that showcase accurate computation using input data. The results are sent to the Sepolia testnet through smart contracts after completing the necessary proofs. These smart contracts validate the proofs and update the state of the blockchain. Once the results have been verified, they can be easily accessed in other smart contracts, ensuring secure and reliable interactions.

\section{Experimental Setup}
The experimental setup used in our study consists of two phases: the proof generation phase and the proof verification phase. The proof generation phase involves an in-depth exploration of the processing environment and configuration details pertinent to a personalized AI model's zk proof generation process. The proof verification phase delves into the implementation steps associated with deploying zero-knowledge proof on the blockchain and verifying zero-knowledge proofs using Chainlink oracles.

\subsection{Proof Generation Phase}
The proof generation setup uses an NVIDIA Jetson TX2, a cutting-edge device known for its high computational power and energy efficiency. The specifications of NVIDIA are listed in Table 1.

\begin{table}[htbp!]
\centering
\begin{tabular}{|>{\raggedright\arraybackslash}p{3cm}|>{\raggedright\arraybackslash}p{4cm}|}
\hline
\textbf{Component} & \textbf{Details} \\ \hline
CPU & 6 ARM Cortex-A57 \\ \hline
GPU & 256-core NVIDIA Pascal  \\ \hline
Memory & 8GB LPDDR4  \\ \hline
\end{tabular}
\caption{Nvidia Jetson TX2 specifications}
\end{table}

We selected this device due to its suitability for AI applications, which are known to require significant computational resources. Our objective was to develop zk-SNARK circuits designed to generate zero-knowledge proofs. These circuits are specifically tailored for a linear regression model, utilizing characteristics obtained from the on-chain data of Bitcoin as a CSV file. The linear regression model coefficients, including the model weights, were saved in a JSON file. We used Python scripts to automate the process of generating circuit files. These scripts received the JSON data and produced multiple Circom files, each representing a distinct number of weights.

Creating and confirming proofs involves building zk circuits using the Circom programming language, generating witnesses, and then proving and checking the proofs using the Snarkjs library. The automated script managed the complete procedure, encompassing compilation, witness production, contribution to the ceremony, preparation for phase 2, zkey generation, and proof generation and verification. We used the Circom tool to generate a smart contract-based verifier that allows proofs to be verified on the blockchain. Remix was used to deploy the Verifer smart contract on the blockchain.  The trusted setup  was conducted by a consortium of stakeholders, including model developers, auditors and decentralized oracle providers. This collaborative approach ensures trust in the setup process and mitigates the risk of a single point of failure.

\subsection{Proof Verification Phase}
For this experiment, we chose the Sepolia testnet because it is widely used among developers and one of the few testnets supported by Chainlink. The experimental findings are relevant and applicable to live production settings like the Ethereum main network. We deployed the verifier smart contract on the testnet for zk verification purposes, ensuring the thoroughness of our testing process.

We set up Chainlink Functions to integrate the decentralized oracle network to the Sepolia testnet. We cloned the Chainlink Functions starter kit from the official GitHub repository \cite{smartcontractkit2024functions}.

This configuration offered the essential resources to interact with the blockchain and Chainlink oracle networks. Subsequently, we modified the Functions request configuration file to explicitly define the source code for API calls and perform computations based on the smart contract request. We established the environment variables using encrypted data for access. This process involved establishing the environment variable file's password and configuring the environment variable by specifying the key and value. We used four keys to setup the experiment:
\begin{itemize}
    \item A private key obtained from the MetaMask wallet.
    \item An Remote Procedure Call (RPC) URL derived from the Alchemy website for the Sepolia testnet.
    \item An API token for GitHub.
    \item An API for the blockchain explorer Etherscan\end{itemize}

\begin{figure}[h!]
    \centering
    \includegraphics[width=1\linewidth]{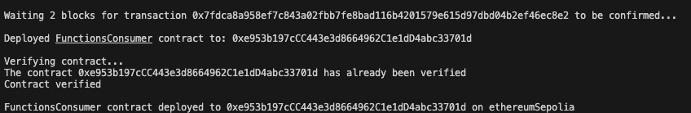}
    \caption{Oracle functions consumer contract deployed to Sepolia.}
    \label{fig:4}
\end{figure}

Upon configuring the environment variables, the functions consumer contract was successfully deployed to the Sepolia testnet, as shown in Fig.~\ref{fig:4}, completing the integration with Chainlink oracles. 

\begin{figure}
    \centering
    \includegraphics[width=1\linewidth]{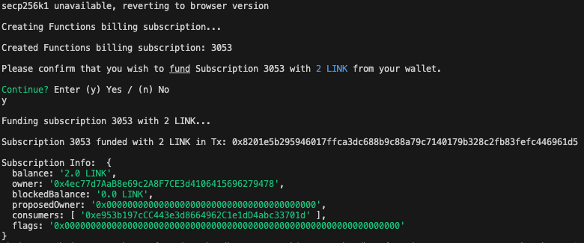}
    \caption{Funding the subscription}
    \label{fig:5}
\end{figure}

The consumer contract address is used to create and fund the billing subscription for Chainlink Functions, as shown in Fig. \ref{fig:5} using LINK tokens acquired via the Chainlink Faucet.

\begin{figure}
    \centering
    \includegraphics[width=1\linewidth]{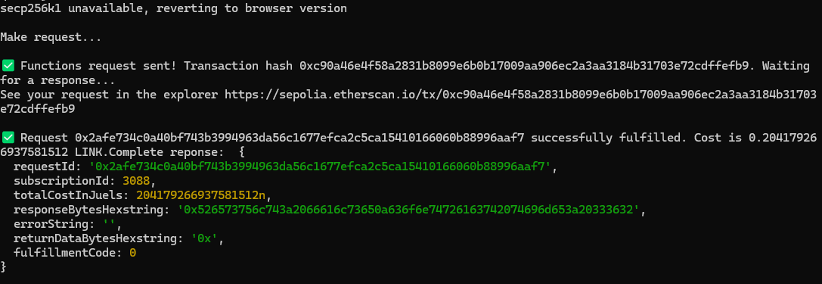}
    \caption{Chainlink functions API and computation Output}
    \label{fig:6}
\end{figure}

The Chainlink's smart contract requests the nodes to perform zk computations and return the result. The proof size of our model is 806 bytes and the verification key size is 2922 bytes. The script runs the functions in a sandbox environment, as seen in Fig.~\ref{fig:5} before making an on-chain transaction to ensure they are correctly configured and the fulfilment costs are estimated before making the request. As shown in Fig.~\ref{fig:6}, chain data retrieval was implemented by pushing API queries to external API providers for on-chain data utilizing the Chainlink Functions. 

\section{Experimental Results and Analysis}

\begin{figure}
    \centering
    \includegraphics[width=1\linewidth]{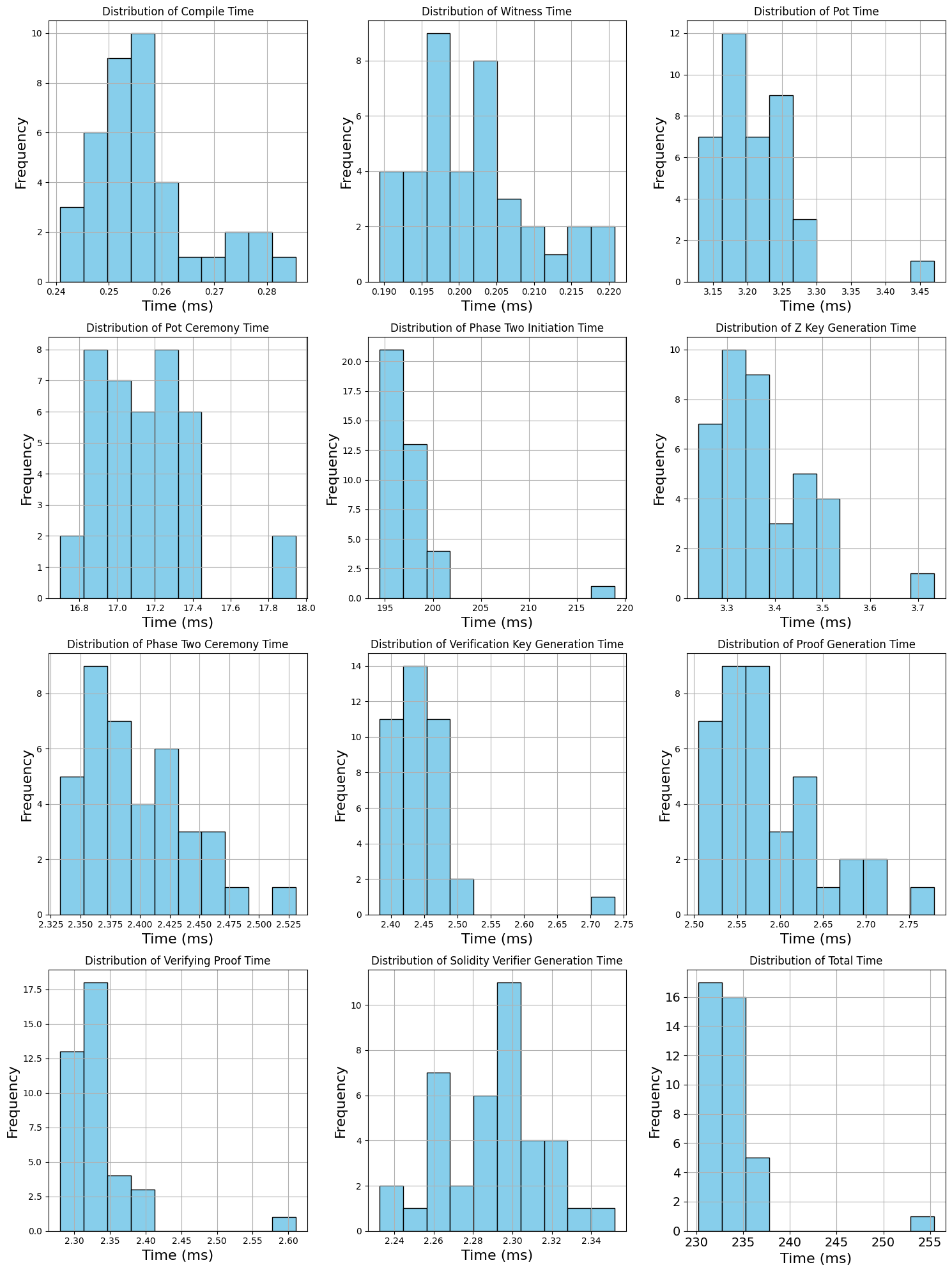}
    \caption{Distribution analysis for each phase of zk generation.}
    \label{fig:7}
\end{figure}

\begin{figure}
    \centering
    \includegraphics[width=1\linewidth]{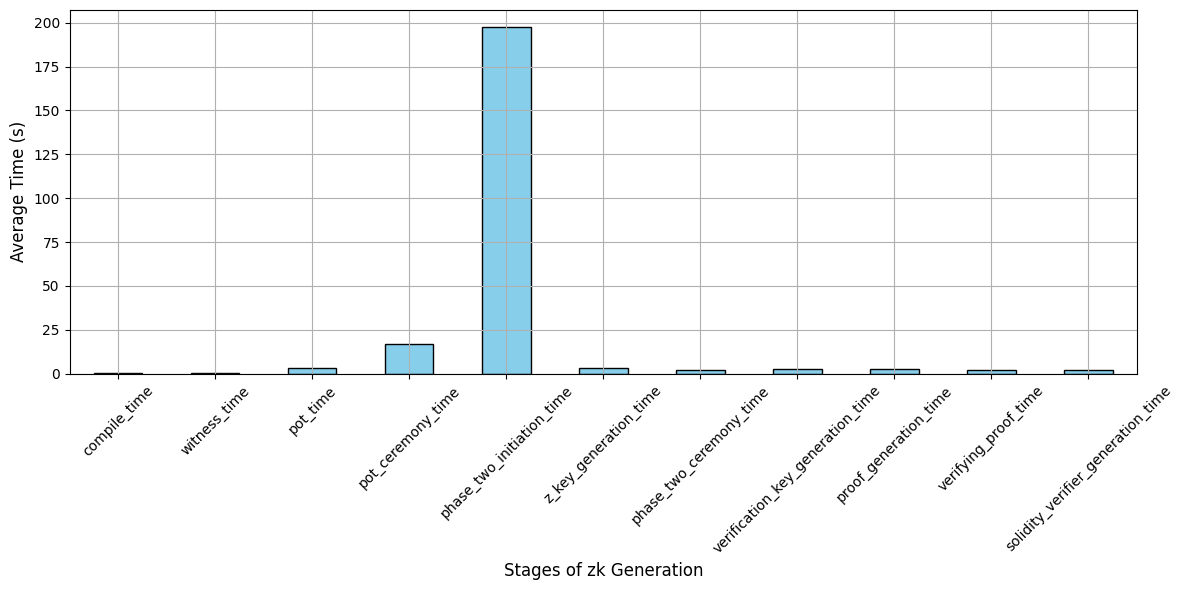}
    \caption{Average time taken for each stage of zk generation.}
    \label{fig:8}
\end{figure}

Our experimental setup aimed to replicate real-life scenarios for deploying and verifying personalized AI models in a blockchain-enabled AI marketplace. Our study is the first to utilize the Chainlink oracle network to compute and evaluate the efficiency of the zk verification for personalized AI models. We used NVIDIA Jetson TX2 to simulate the developer's process of generating zk-SNARK proofs for their trained personalized AI models before deployment. The zk-SNARK verification was conducted on the Sepolia testnet, and the zk verification computations were performed using Chainlink oracles to ensure secure and reliable verification.

We assess the efficiency and resource consumption of the zk-SNARK generation and verification process for personalized AI models. We also evaluate the overheads introduced by blockchain and Chainlink oracles during the verification process. This study aims to demonstrate the feasibility and effectiveness of using zk-SNARKs and Chainlink oracles to verify personalized AI models securely and efficiently.

The time analysis of zk-SNARK proof generation and verification involved examining various stages as shown in Fig. \ref{fig:7} and Fig. \ref{fig:8} and focusing on their duration and variability using a distribution analysis. The compilation process is the process of converting the linear regression model into an arithmetic circuit for zk-SNARK proof generation. The average time to compile our model is efficient and took approximately 0.256 seconds. Witness time involves creating the internal model values required for zk-SNARK proof generation. This will be used as cryptographic evidence to show the validity of the computations without revealing inputs. The witness time distribution shows low mean value of 0.202 seconds. The Power of Tau (PoT) which is a crucial phase in the zk-SNARK trusted setup process takes an average of 3.21 seconds. During this phase, cryptographic parameters are generated to ensure   reliability of the zk-SNARK system, allowing it to produce proofs without revealing private information. The PoT ceremony time process involves multiple participants to contribute randomness to generate the final parameters taking 17.14 seconds. These resulting parameters are known as the common reference string (crs) and are necessary for any zk-SNARK proofs generated by the system. 

Phase two initiation time was the most computationally demanding phase taking approximately 197.39 seconds as shown in Fig. \ref{fig:8}, this is due to the complex setup of cryptographic parameters for zk-SNARKs. The time required for this stage is heavily contingent on the size and complexity of the personalized AI model in our case a linear regression model being converted into an arithmetic circuit for zk-SNARK proof generation. The complexity of QAP constraints increases with more complex AI models as they have larger number of features. This is also evident in the wide distribution of phase two initiation time, indicating significant differences in processing times adding to the longer proof generation times. The generation of the zk key takes 3.37 seconds indicating that it is relatively efficient once the cryptographic setup is completed. The verification and proof generation times are much faster than earlier stages like PoT and phase two initiation taking 2.59 and 2.45 seconds. This is due to the nature of zk-SNARKs producing succinct cryptographic proofs allowing for quick proof generation and verification irrespective of the complexity of AI models.

\begin{figure}
    \centering
    \includegraphics[width=1\linewidth]{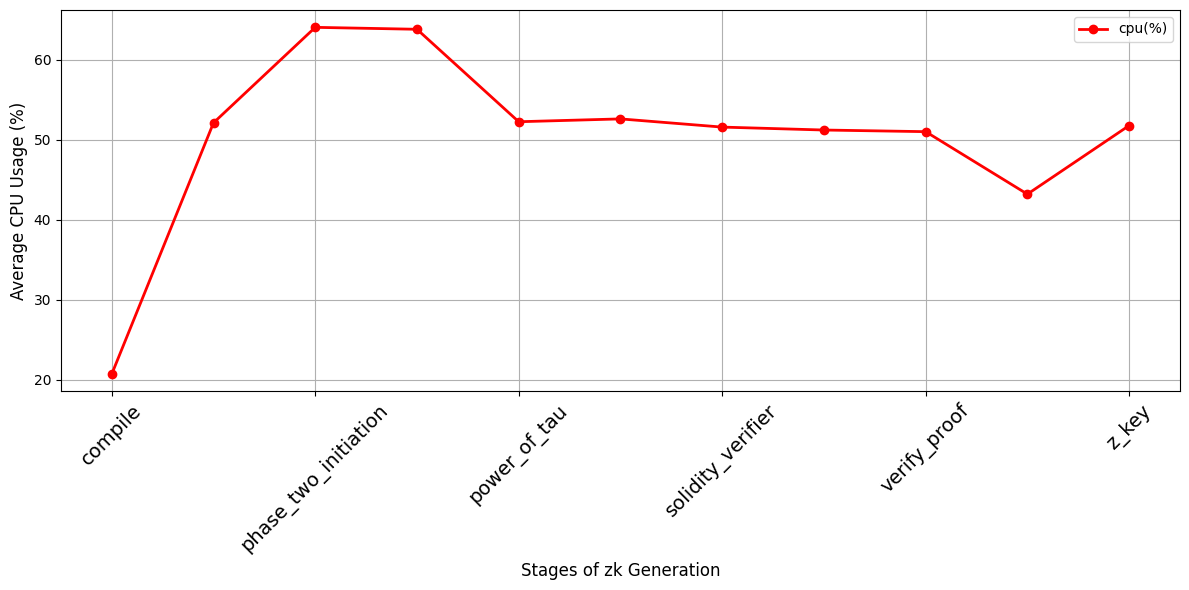}
    \caption{Average CPU usage for each stage of zk generation.}
    \label{fig:9}
\end{figure}

\begin{figure}
    \centering
    \includegraphics[width=1\linewidth]{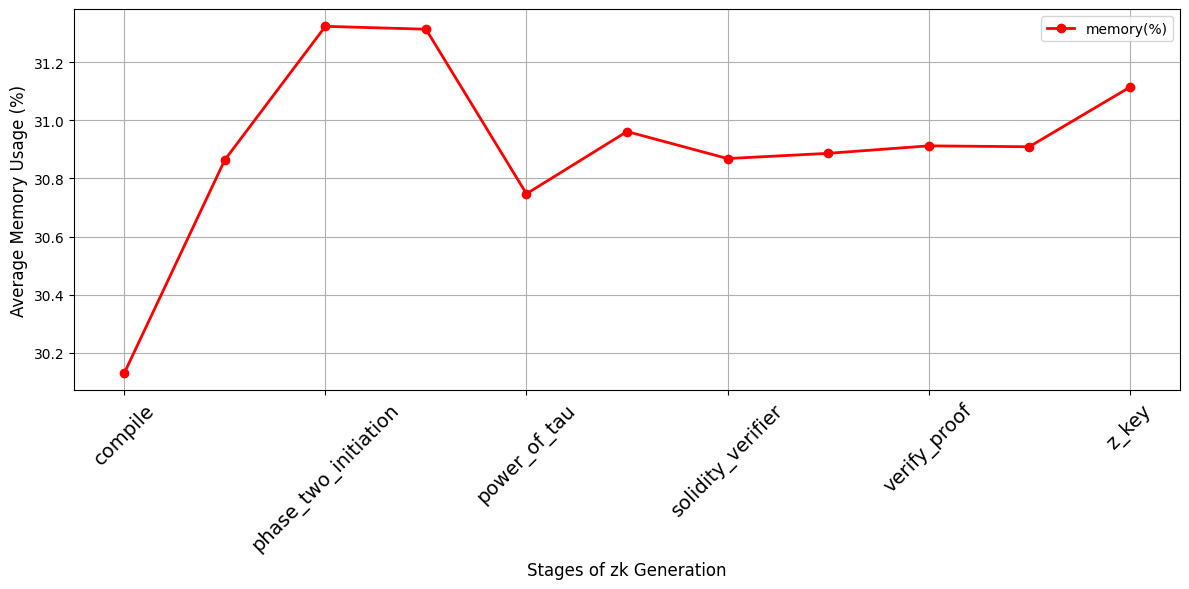}
    \caption{Average RAM usage for each stage of zk generation.}
    \label{fig:10}
\end{figure}

Looking at the average time for each stage reflected in Fig.~\ref{fig:8}, it is evident that phase two initiation time stood out as the most time-consuming stage, followed by the power of tau (PoT) and the phase two ceremony. In comparison, compile, witness, and zk key generation times are notably shorter. Overall, zk Proof Generation takes significantly longer, averaging 233.63 seconds, compared to zk Verification, which took 61.50 seconds. We analyzed CPU and memory consumption to understand the resource requirements encountered during the various phases of zk-SNARK proof creation. From Fig.~\ref{fig:9}, we can see that CPU usage was highest during the Phase Two Initiation and Power of Tau stages, indicating these stages are particularly computationally intensive. Other stages like compile, proof, and verification key generation also showed significant CPU usage but to a lesser extent. Memory usage remained relatively consistent across all stages as seen in Fig.~\ref{fig:10} hovering around 30-31\%, with slight variations observed during the Phase Two Initiation and Power of Tau stages, which can be attributed to the intensive computations required for phase two and power of tau setup. This indicates that developers will face penalties for higher resource consumption and longer times during the zk-SNARK proof generation phase, especially if their models are complex or inefficient. Therefore, optimizing the proof generation process is crucial to avoid high computational costs and delays.

\begin{figure}
    \centering
    \includegraphics[width=0.7\linewidth]{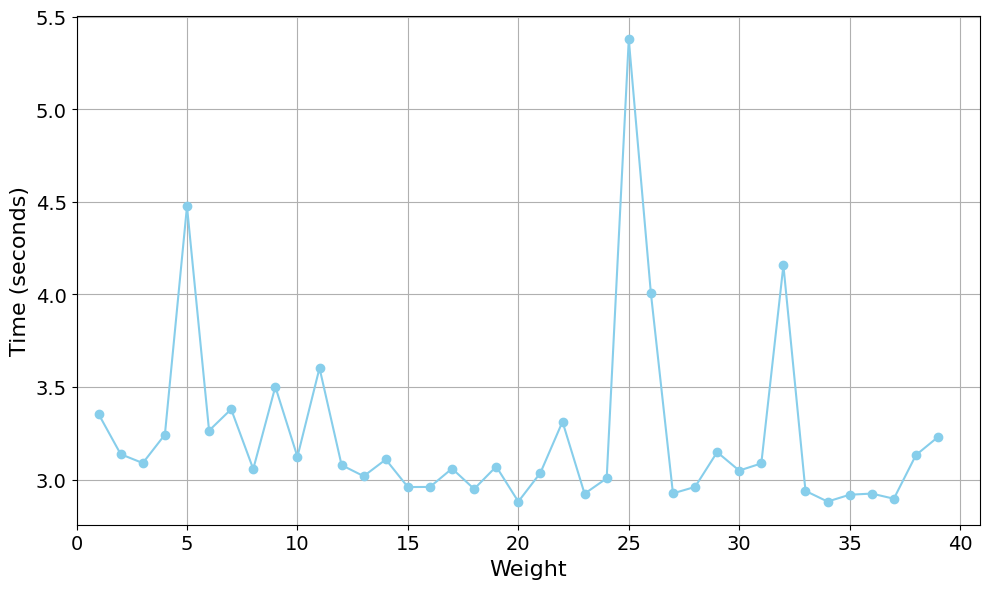}
    \caption{Blockchain and Chainlink overhead time over 39 weights.}
    \label{fig:11}
\end{figure}

\begin{figure}
    \centering
    \includegraphics[width=1\linewidth]{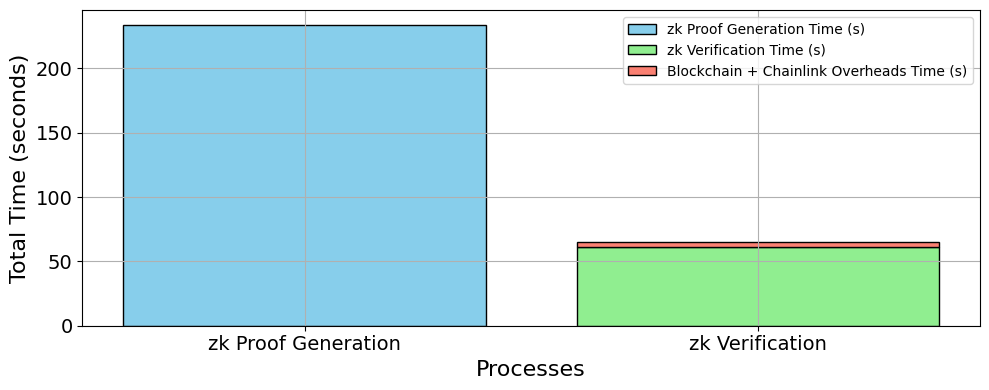}
    \caption{Comparison of time taken for zk proof generation vs zk verification process.}
    \label{fig:12}
\end{figure}

Figures \ref{fig:11} and \ref{fig:12} show that blockchain and Chainlink oracle overhead times were minimal compared to the zk proof generation and verification times, highlighting the efficiency of using Chainlink oracles for decentralized verification. Zk-SNARKs are designed to provide a compact proof that can be verified quickly, regardless of the underlying complexity of the original computation. Chainlink oracle network was utilized to compute zk-SNARK verification and return the result to the blockchain, ensuring that the process is both efficient and secure. Figure~\ref{fig:12} demonstrates that the zk verification process is efficient. The users can be assured that AI models are verified securely and efficiently, as Chainlink's decentralized oracle network adds an extra layer of robustness by eliminating single points of failure. This decentralized verification process ensures that the zk-SNARK proofs are validated in a trust-minimized manner.

\begin{figure}
    \centering
    \includegraphics[width=1\linewidth]{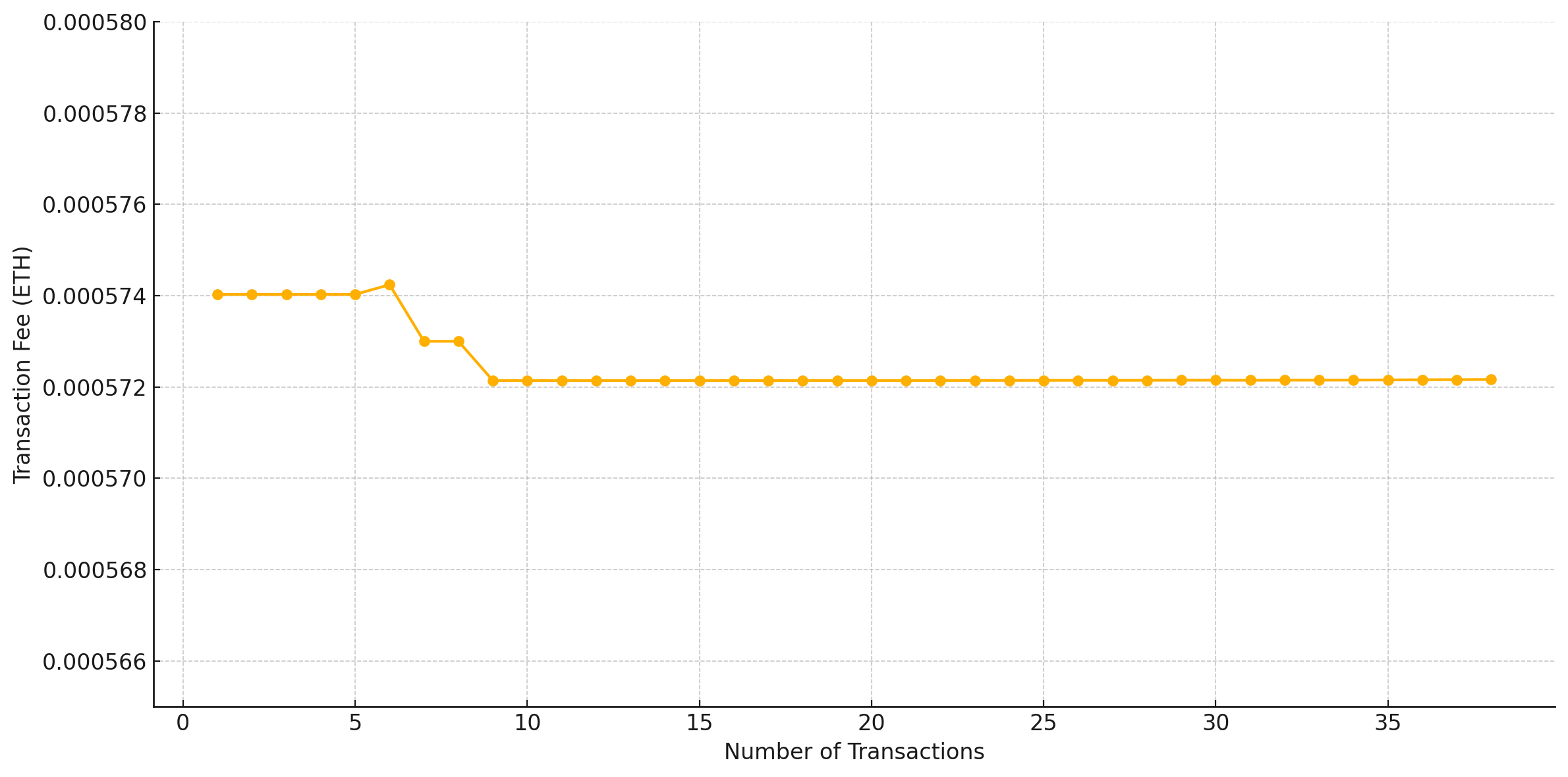}
    \caption{Transaction Fee (ETH) for zk verification.}
    \label{fig:13}
\end{figure}

The results in Fig. \ref{fig:13} indicate the transaction fees associated with the zk-SNARK verification requests on the Ethereum blockchain. The dataset comprised 39 transactions, each representing a distinct zk verification request. With an average fee of 0.000572 ETH, translating to \$1.03 USD for each verification. The transaction fees were relatively consistent across all transactions, with minimal variability.

\begin{figure}[h!]
    \centering
    \includegraphics[width=1\linewidth]{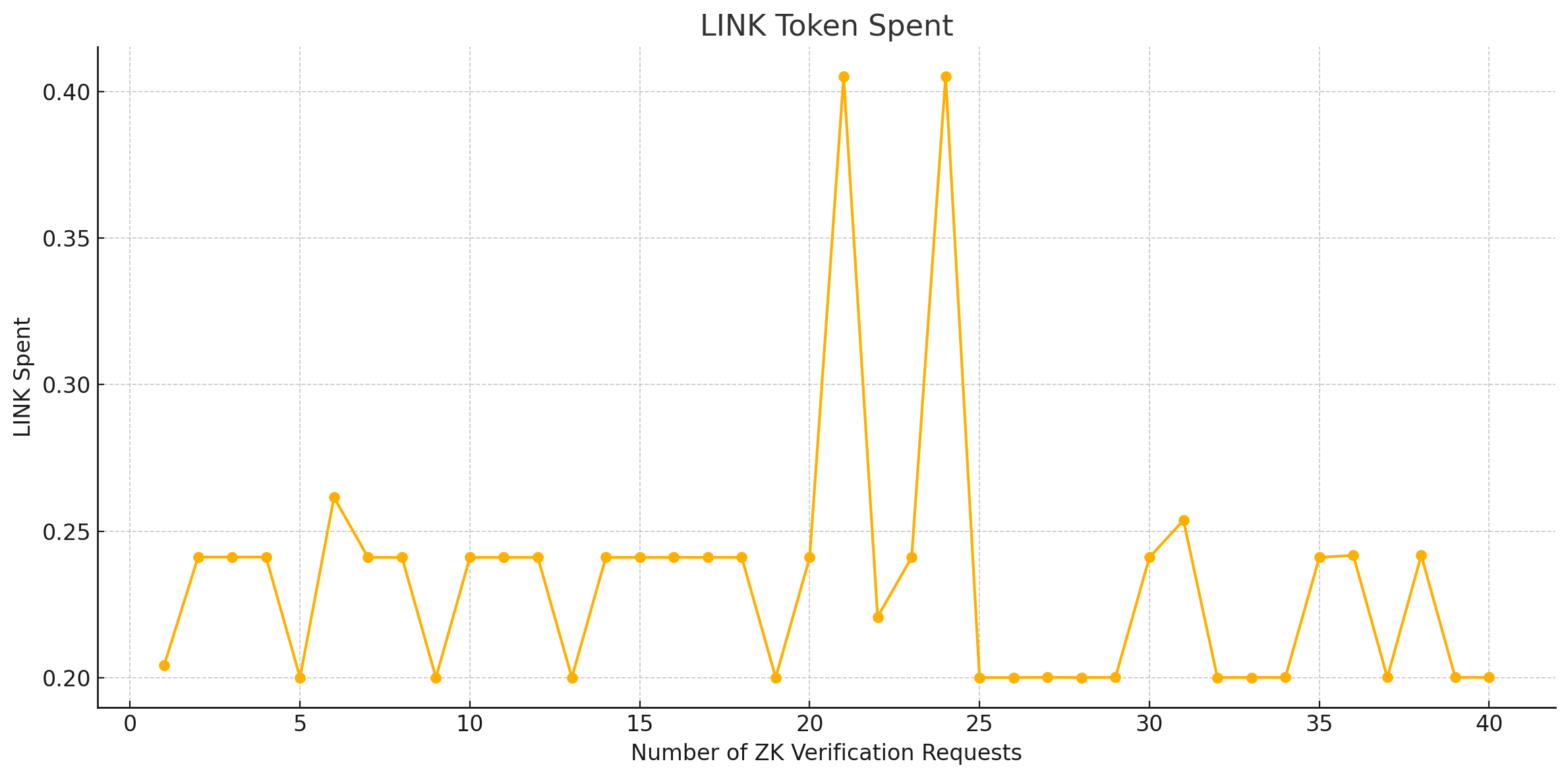}
    \caption{Amount of LINK token spent for zk verification.}
    \label{fig:14}
\end{figure}

Analyzing transaction fees and LINK token expenditure provide valuable insights into the cost structure of deploying zk-SNARK verifications in a decentralized environment. The consistent transaction fees suggest a predictable cost model, benefiting developers and operators planning to integrate such verifications into their systems.  At the time of this writing, the cost of LINK is approximately \$14.16 USD per token \cite{chainlink2024}, this translates to a range of \$2.83 USD to \$5.66 USD per request for a zk verification computation. As seen in Fig.~\ref{fig:14}, the LINK token expenditure graph further emphasis that the costs associated with chainlink oracle are stable for the verification process.

The transaction fees were plotted against the number of transactions to visualize the fee distribution. In addition to transaction fees, we analyzed the LINK token expenditure for the oracle requests involved in zk-SNARK verification. The analysis included 39 oracle requests for zk verification computations, and the expenditure was plotted against the number of zk verification requests. This graph helped identify the cost distribution across different requests, highlighting any peaks that might indicate higher computational or operational demands. Previous work used selective verification to reduce the number of verifications and hence the overall costs associated \cite{south}. While costs associated are higher than centralized systems due to the decentralized nature of blockchain and oracles, these costs are justified by the added benefits of increased trust and continuous transparency for verification. 

\section{Discussion}
Our analysis in the results section indicates that the speed of proof generation is the main constraint that requires significant resources and is process intensive. This can be attributed to the complexity of creating a QAP from the arithmetic circuit, which introduces arithmetization constraints that are difficult to address. These constraints ensure an accurate polynomial representation of AI model operations, but their complexity increases with model size and feature count. For models more complex than a linear regression model, such as a deep learning network, the number of required gates and constraints can increase exponentially, leading to significant resource consumption, longer proof generation times and could present scalability issues. While our framework shows the feasibility of using zk-SNARK-based verification for a linear regression model on blockchain, further optimization is necessary to improve efficiency of zk-SNARK proof generation enabling the use of more advanced AI models within this framework. Techniques such as proof splitting \cite{Qi2023SplitAH}, GPU acceleration \cite{Derei2023AcceleratingTP}, and parallel processing \cite{Lu2023cuZKAZ} of zk-SNARK proofs using tools such as Sonic \cite{Maller2019SonicZS} have shown promise in improving the efficiency and reducing costs for zk-SNARK proof generation.

\begin{table*}[h!]
\centering
\renewcommand{\arraystretch}{1.3} % Increase row height
\large % Increase text size
\label{tab:comparison}
\resizebox{\textwidth}{!}{%
\begin{tabular}{|>{\centering\arraybackslash}m{3cm}|>{\centering\arraybackslash}m{4.5cm}|>{\centering\arraybackslash}m{4cm}|>{\centering\arraybackslash}m{4cm}|>{\centering\arraybackslash}m{4cm}|>{\centering\arraybackslash}m{4cm}|>{\centering\arraybackslash}m{4cm}|}
\hline
\textbf{Key Attributes}           & \textbf{Our Paper} & \textbf{SNNzksNARK \cite{SNNzksNARK}} & \textbf{ Verifiable Evaluations of ML using zk-SNARKs \cite{south}} & \textbf{Trustless DNN Inference \cite{Kang2022ScalingUT}} &  \textbf{zkCNN \cite{Liu2021zkCNNZK}} & \textbf{Secure Machine Learning using Homomorphic Encryption \& Verifiable Computing \cite{HEVC}} \\ 
\hline
\textbf{Blockchain \& Oracle Integration}
& Yes & No & No & No & No & No \\
\hline
\textbf{Decentralization}    & Fully decentralized using blockchain and Chainlink decentralized oracle network (DON) & Centralized & Centralized & Centralized; runs inference verification in a centralized MLaaS (Machine Learning-as-a-Service) model &  Centralized & Centralized \\ \hline
\textbf{Trust}               & Completely trustless; leverages blockchain, zk-SNARKs and Chainlink oracles for verifiable computations & Relies on trusted central entities for computation and validation & Partially trustless; ensures correct model inference but still requires trust in model providers not to swap models & Limited trustless; relies on a trusted ML service provider to generate zk-SNARKs proofs & Limited trustless;  Relies on the model provider to generate and distribute proofs honestly & Partially trustless; relies on a centralized entity to manage HE-encrypted data\\ \hline
\textbf{Transparency}        & Fully transparent due to Chainlink oracles and zk-SNARK integration; verifications are stored on-chain & Limited transparency as proofs are stored on a centralized setup; lacks public auditability & Provides transparency in verifiable inference but lacks blockchain immutability; results are not recorded on a auditable ledger & Limited transparency as proofs are not public and depends on centralized storage &  Limited Transparency; proofs ensure inference correctness, but they are not publicly auditable  & Limited Transparency; does not store the proofs on a publicly verifiable ledger \\ \hline
\textbf{Privacy}             & Ensures privacy by integrating zk-SNARKs into Chainlink oracles for proof verification & Privacy is protected using zk-SNARKs but trained neural network weights might still be exposed & Strong focus on ML model privacy using zk-SNARKs & Ensures input and model privacy via zk-SNARKs but data exposure risks exist & zkCNN hides CNN weights and input data, ensuring confidential inference verification & Ensures privacy of model inputs and outputs using homomorphic encryption (HE) \\ \hline

\textbf{Evaluation Methodology} & Real-world implementation using the NVIDIA Jetson TX2, Sepolia testnet and Chainlink's DON & Simulated performance evaluation on neural networks; No real-world deployment & Real-world tests using actual ML models and benchmarked proof generation & Simulations for centralized environments; no real-world test & zkCNN is benchmark-tested, lacks practical real-world implementation & Benchmarked three architectures for ML evaluation\\ \hline

\textbf{AI/ML Model Verification} & Yes - zk-SNARK proofs verify AI model performance claims without revealing model weights & Yes - Uses zk-SNARKs for verifying neural network execution & Yes, verifies inference correctness using zk-SNARKs without revealing model weights & Yes - Focuses on verifying AI inference correctness & Yes, zkCNN guarantees correct CNN  model inference execution & No AI-specific verification \\ \hline
\textbf{Data Verification}   & Yes - Chainlink oracles fetch and verify off-chain data before model verification & No - explicit data verification mechanism; assumes correct data input & No - Does not use external verification mechanisms; assumes data inputs are correct & No - Does not use external verification mechanisms; assumes data inputs are correct & No - assumes data correctness without independent verification & No external data verification \\ \hline
\textbf{Scalability}         & Scales efficiently due to use of Chainlink oracles for zk-SNARK computations & More efficient scaling due to centralized environments & Limited scalability; real-time AI model inference verification is computationally expensive & Scales better in centralized MLaaS setups & Highly scalable for CNN-based verifications, not designed for large-scale AI verification & Limited scalability due to high computational costs of homomorphic encryption\\ \hline
\textbf{Efficiency}          &  Comparatively higher computational overhead due to decentralized architecture & More efficient; Highly optimized for single-server performance due to centralized architecture with optimized environments &  More efficient than decentralized solutions but less efficient than centralized solutions due to the computationally intensive privacy-preserving inference proofs & More efficient; Optimized for centralized systems with batched proofs & Highly efficient for CNN inferences & Computationally expensive due to homomorphic encryption and verifiable computing overhead \\ \hline
\end{tabular}%
}
\caption{\centering Comparison with the existing  verification methods across key attributes.}
\end{table*}

To our knowledge, none of the above studies has implemented zk-SNARKs on a practical blockchain system and a decentralized oracle network to verify AI models. Direct comparisons to existing non-blockchain zk-SNARK AI verification implementations, such as those in \cite{SNNzksNARK}, \cite{south}, \cite{Kang2022ScalingUT}, and \cite{Liu2021zkCNNZK}, are challenging due to differences in the underlying systems and AI models tested. While these studies focus on implementing zk-SNARKs in centralized systems, our work integrates zk-SNARKs into a decentralized blockchain and oracle network. Despite the inherent differences in our blockchain-based implementation compared to centralized systems, we can still draw important conclusions based on our results. 

Existing verification methods such as HE and VC \cite{HEVC} and implementations of zk-SNARKs such as those in \cite{SNNzksNARK}, \cite{south}, \cite{Kang2022ScalingUT} and \cite{Liu2021zkCNNZK}, benefit from optimized environments where data and computational resources are centrally managed. These setups enable faster proof generation and verification by reducing communication overhead and leveraging high-performance infrastructure, such as dedicated servers or centralized cloud systems. The key trade-off in implementing zk-SNARKs in decentralization systems is increased transparency and trust at the cost of increased transaction fees and efficiency compared to centralized systems. Decentralized oracles take longer to fetch and verify data, compute the zk-SNARK proof and the blockchain verification adds further delays due to the decentralized nature of the network both of which slow down the process. While this ensures trustless, transparent verification, it results in reduced efficiency compared to zk-SNARKs implementations on centralized systems for AI verification. The key differences in attributes between our approach and existing verification methods is highlighted in Table 2.

%The key trade-off in implementing zk-SNARKs in decentralizated systems is increased transparency and trust at the cost of scalability, increased transaction fees and efficiency compared to centralized systems. Decentralized oracles take longer to fetch and verify data, compute the zk-SNARK proof and the blockchain verification adds further delays due to the decentralized nature of the network both of which slow down the process. While this ensures trustless, transparent verification, it results in slower performance, increased costs and reduced scalability compared to zk-SNARKs implementations on centralized systems for AI verification. 

A significant incentive for participants to engage with this framework lies in the model data privacy and trustless verification offered by zk-SNARKs, especially when used in conjunction with blockchain technology and decentralized oracles. The decentralized nature of blockchain and oracles ensures that no single entity controls the data or verification process, enhancing transparency and preventing tampering with transaction records. For developers, the ability to verify AI model performance without exposing proprietary data such as model weights ensures that their intellectual property remains secure, reducing the risk of misuse and unauthorized replication. These guarantees encourage developers to bring innovative AI models to the marketplace with confidence knowing that their investments are safeguarded in a trustless and immutable environment. For buyers, zk-SNARKs and the decentralized infrastructure offer a reliable means to independently verify claims of AI models, ensuring that the claims made by sellers about model performance are accurate and trustworthy. This capability promotes transparency and trust in AI models in blockchain-enabled marketplaces, enabling buyers to make informed decisions based on verifiable evidence of model efficacy, thereby fostering a more trustworthy and equitable ecosystem.

%put in the argument of how using zk-snarks in blockchain systems means trade off compaared to other centralized studies. 

\section{Conclusion}

This paper presents a novel framework for verifying AI model performance claims on blockchain. Our study indicates that the zk proof generation process is the most time-consuming and computationally intensive stage. Optimizing this stage is crucial for enhancing the overall efficiency of zk-SNARK implementations. The zk Verification process on Chainlink oracles is relatively faster but still significant compared to the overhead time, emphasizing the importance of efficient verification mechanisms.

By using the NVIDIA Jetson TX2 for local proof generation and the Sepolia testnet with Chainlink oracles for decentralized verification, our study demonstrates a robust and feasible approach to securely and transparently verifying personalized AI models using a real-world oracle network and testnet setup. Integrating Chainlink oracles with the Sepolia testnet environment allowed us to replicate real-world conditions, providing insights into the practical challenges and benefits of deploying zk-SNARKs in decentralized settings. This implementation highlights the feasibility of using Chainlink's decentralized oracle network to handle the computational demands of zk-SNARK verification in real-world applications. The consistent performance and minimal overhead observed during our tests indicate that such a setup can effectively manage the verification of personalized AI models at scale. Furthermore, the scalability of Chainlink oracles ensures that this approach can accommodate increasing verification demands without compromising efficiency.

Our findings highlight the potential of combining zk-SNARKs with decentralized oracle networks to improve the transparency and privacy of AI model verification processes for blockchain in real-world applications.
In addition to showing that this framework is technically feasible, this study lays the groundwork for future studies that optimize zk-SNARK proof generation and investigate broader applications of Chainlink oracles for AI verification on blockchain. 
In future work, we will conduct a comprehensive security evaluation of the proposed framework to address potential vulnerabilities and evaluate its robustness against various attack scenarios. This will ensure that the framework is not only efficient and scalable but also secure, thus increasing its applicability and trustworthiness in blockchain-enabled AI marketplaces.
\sloppy
\bibliographystyle{IEEEtran} 
\bibliography{Preprint}

\vskip -2\baselineskip plus -1fil

%---------------------------------------------------------
% Some main sections of your paper would go here ...
% e.g. \section{Introduction}, \section{Methodology}, etc.
%---------------------------------------------------------

\end{document}